\documentclass[sn-mathphys,Numbered,iicol]{sn-jnl}

\usepackage{graphicx}%
\usepackage{multirow}%
\usepackage{amsmath,amssymb,amsfonts}%
\usepackage{amsthm}%
\usepackage{mathrsfs}%
\usepackage[title]{appendix}%
\usepackage{xcolor}%
\usepackage{textcomp}%
\usepackage{manyfoot}%
\usepackage{booktabs}%
\usepackage{algorithm}%
\usepackage{algorithmicx}%
\usepackage{algpseudocode}%
\usepackage{listings}%

\begin{document}

\title{Epitaxy enhancement in oxide/tungsten heterostructures by harnessing the interface adhesion}

\author*[1]{Anna L. Ravensburg}\email{anna.ravensburg@physics.uu.se}

\author[2]{Rimantas Brucas}

\author[3,4]{Denis Music}

\author[1]{Lennart Spode}

\author[1]{Gunnar K. P\'{a}lsson}

\author[2]{Peter Svedlindh}

\author[1]{Vassilios Kapaklis}

\affil*[1]{\orgdiv{Department of Physics and Astronomy}, \orgname{Uppsala University}, \orgaddress{\street{Box 516}, \city{Uppsala}, \postcode{75120}, \country{Sweden}}}

\affil[2]{\orgdiv{Department of Materials Science and Engineering}, \orgname{Uppsala University}, \orgaddress{\street{Box 516}, \city{Uppsala}, \postcode{75120}, \country{Sweden}}}

\affil[3]{\orgdiv{Department of Materials Science and Applied Mathematics}, \orgname{Malm{\"o} University}, \orgaddress{\city{Malm{\"o}}, \postcode{20506}, \country{Sweden}}}

\affil[4]{\orgdiv{Biofilms Research Center for Biointerfaces}, \orgname{Malm{\"o} University}, \orgaddress{\city{Malm{\"o}}, \postcode{20506}, \country{Sweden}}}


\abstract{The conditions whereby epitaxy is achieved are commonly believed to be mostly governed by misfit strain. We report on a systematic investigation of growth and interface structure of single crystalline tungsten thin films on two different metal oxide substrates, Al$_{2}$O$_{3}$~($11\bar{2}0$) and MgO~($001$). We demonstrate that despite a significant mismatch, enhanced crystal quality is observed for tungsten grown on the sapphire substrates. This is promoted by stronger adhesion and chemical bonding with sapphire compared to magnesium oxide, along with the restructuring of the tungsten layers close to the interface. The latter is supported by {\it ab initio} calculations using density functional theory. Finally, we demonstrate the growth of magnetic heterostructures consisting of high-quality tungsten layers in combination with ferromagnetic CoFe layers, which are relevant for spintronic applications.}


\maketitle

\section{Introduction}\label{sec1}

Spintronic devices, consisting of ferromagnetic layers separated by a nonmagnetic metal or an insulating layer, use spin-dependent electron transport to detect changes in magnetic fields \cite{Spintronics_Review_Hillebrands_2020}. In light of this, heterostructures of ferromagnetic layers in proximity to 4$d$ and 5$d$ non-magnetic metals are of particular interest. For example, such heterostructures can be used to tune the strength and type of interlayer exchange coupling in trilayers and to fine-tune the magnetization dynamics \cite{Parkin_RKKY_1991, Liu_CoFeV_Ru_JAP_2022}. Most of these heterostructures have to be grown on oxide substrates, making the oxide/metal interface with its chemistry and structure on multiple length scales an important parameter to consider while designing and evaluating the performance of a device.

Spin-orbit torques (SOTs) in heavy metal/ferromagnetic heterostructures are gaining increasing attention for providing an efficient pathway for manipulating the free layer magnetization in magnetic random-access memories. The origin of the SOT is the pure spin current $J_s$ generated by a charge current $J_c$ in the heavy metal via the spin-orbit coupling. The charge-to-spin current conversion efficiency can be described by the spin Hall angle $\theta_{SHE} = J_s/J_c$ \cite{SH_W_APL_2012}. Tungsten has, in this respect, been in focus due to large reported values of $\theta_{SHE}$ of around $-0.3$ to $-0.4$. An equally important parameter for spintronic applications is the transparency of the heavy metal/ferromagnet interface which is controlled by the effective spin-mixing conductance of the interface \cite{Tserkovnyak2002}. The effective spin-mixing conductance describes the transfer of spin current through the interface and also accounts for spin-backflow as well as spin-memory loss at the interface, emphasizing that the interface quality is a critical parameter for spintronic devices. Furthermore, W has been the subject of investigations as fusion reactor plasma-facing material due to its combination of high atomic number and relatively low activation decay time \cite{Causey2001, Kaufmann2007}. For these applications, being able to grow well defined crystals of these metals and performing interface engineering is of the outmost importance; in particular for a potential new generation of spintronic devices where inter-layer spin transport play an important role.

In thin-film form, besides the ground state bcc $\alpha$-W phase, tungsten can be stabilized in its $\beta$-W phase with an A15 structure \cite{Keefe1996, betaW_PRA}. The $\beta$-W phase exhibits a large spin Hall angle and therefore also a large charge-to-spin current conversion efficiency, similar to the case of $\beta$-Ta \cite{SH_W_APL_2012, betaW_PRA, betaW_APL, betaW_SR}. The $\beta$-W phase growth depends strongly on the deposition conditions during the sputtering process as well as on the film thickness \cite{Shen2000, betaW_APL, VULLERS201526}. Thin films may exhibit the $\beta$-W phase, while intermediate thicknesses and/or annealing yield mixture of $\alpha$-W and $\beta$-W phase \cite{Rossnagel2002, Keefe1996}. Thick films tend to be almost pure $\alpha$-W phase. Furthermore, the majority of these films are polycrystalline forming more complicated interfaces at grain boundaries, with the adjacent substrates, and with additional buffer-layers. Hetero-epitaxial growth of $\alpha$-W thin films on the other hand, may be enabled through a lattice match between in-plane atomic distances in the film and substrate. The crystal structure of bcc $\alpha$-W is reported to have a cubic lattice parameter of 3.155~{\AA} \cite{Wheeler1925} to 3.17~{\AA} \cite{Jain2013, osti_W}.

Here, we study the sputter growth of highly epitaxial $\alpha$-W thin films on sapphire, Al$_{2}$O$_{3}$~($11\bar{2}0$), and MgO~($001$) substrates, giving emphasis to the overall quality of the layering and crystal structure. Magnetron sputtering was selected for thin film growth because it offers the best compromise between low defect density and flat layering compared to other physical vapor deposition processes. We further shed more light on interdependencies between the oxide/film interface and the accommodation of the lattice mismatch, with support from {\it ab~initio} calculations. The Al$_{2}$O$_{3}$~($11\bar{2}0$) and MgO~($001$) substrates were chosen as they are technologically relevant materials, which are used to grow well defined single crystalline metallic layers and, for the case of W growth, offer very different lattice mismatches, which allows for an in-depth study of the epitaxial growth. Another reason for choosing these substrates is that they provide two different interfaces for tungsten; namely W~($110$) in case of Al$_{2}$O$_{3}$~($11\bar{2}0$) and W~($001$) in case of MgO~($001$). It has been shown that the W~($110$) interface exhibits Dirac-type surface states \cite{Thonig2016}, which is believed to have direct impact on spin Hall angle and therefore also on the generation of spin currents. It is therefore motivated to investigate if the two interfaces are different in this respect. In a forthcoming publication we will present results for the generation of THz radiation in W/CoFe bilayers grown on these two substrates. For the case of sapphire substrates, we argue that the interface structure and crystal quality are results of strong adhesion and bonding, similar to the sapphire/Nb system \cite{Song1997, Wildes2001}. This has as a prerequisite a well-defined epitaxial relationship at the metal/oxide interface and is thus coordination-specific. Having built a solid foundation for growth of epitaxial tungsten, we proceed to the growth of epitaxial $\alpha$-W and ferromagnetic CoFe bilayers. These bilayers are of technological importance, as they might be essential in future spintronic applications, such as THz emitters \cite{THz_review}.

\section{\label{sec:methods}Methods}\label{sec2}

\subsection{\label{sec:growth}Growth}\label{subsec1}

Thin layers of W and W/CoFe bilayers of different thicknesses were deposited on single crystalline Al$_{2}$O$_{3}$~($11\bar{2}0$) and MgO~($001$) substrates (both 10$\times$10~mm$^2$) at floating potential, using direct current (dc) and radio frequency (rf) magnetron sputtering. While for the bilayers, first the W and then the CoFe layer were grown on Al$_{2}$O$_{3}$~($11\bar{2}0$), the order of the layers was turned around for bilayers grown on MgO~($001$) due to better lattice matching of CoFe on MgO. Prior to deposition, the substrates were cleaned in acetone and 2-propanol using ultrasonic agitation for 120~s. This was followed by annealing in vacuum at 873(2)~K for 3600~s. The base pressure of the growth chamber was below 5$\times$10$^{-7}$~Pa. In order to prevent surface oxidation of the films, the samples were capped at ambient temperature ($<$~313(2)~K) with Al; selected samples were capped with Al$_{2}$O$_{3}$ instead. The depositions were carried out in an Ar atmosphere (gas purity $\geq$~99.999~\%, and a secondary getter based purification) from elemental W (25~W, dc) and Al (50~W, dc) targets, and CoFe (13~W, dc) and Al$_{2}$O$_{3}$ (90~W, rf) compound targets. The targets were cleaned by sputtering against closed shutters for at least 60~s prior to each deposition. The target-to-substrate distance in the deposition chamber was around 0.2~m. The deposition rates (W: 0.23~{\AA}/s, Al: 0.30~{\AA}/s, CoFe: 0.10~{\AA}/s, Al$_{2}$O$_{3}$: 0.03~{\AA}/s) were calibrated prior to the growth, using x-ray reflectivity. The growth temperatures for each layer and for the bilayers were optimized with respect to layering and crystal quality, yielding 843(2)~K for single W layers (one selected sample was grown at 793(2)~K instead) and 573(2) K for CoFe layers. For the W/CoFe bilayers, W and CoFe were desposited at 843(2)~K and 573(2)~K, respectively, if W was grown first, while both layers were deposited at 573(2)~K if the CoFe layer was grown first. Finally, in order to ensure thickness uniformity, the substrate holder was rotated during the deposition.

\subsection{\label{sec:characterization}Characterization}\label{subsec2}

X-ray reflectometry (XRR) and diffraction (XRD) were carried out in a Bede D1 diffractometer equipped with a Cu $K_{\alpha_1}$ x-ray source operated at 35~mA and 50~kV. A circular mask (diameter: 0.005~m) and an incidence and a detector slit (both 0.0005~m) were used. For monochromatizing the beam by reducing the CuK$\beta$ and CuK$\alpha_2$ radiation, the setup included a G\"obel mirror and a 2-bounce-crystal on the incidence side. The x-rays were detected with a Bede EDRc x-ray detector. The instrument angles were aligned to the sample surface for XRR and to the W crystal planes for XRD measurements. The measured XRR data was fitted using \textsc{GenX} \cite{Bjorck2007, Glavic_Bjorck_2022} enabling the determination of the scattering length density (SLD) profile which includes information on layer thickness and roughness. However, atomic terraces in the Al$_{2}$O$_{3}$ \cite{Yoshimoto1995} and twinning in the MgO substrates \cite{Schroeder2015}, and therefore also in the epitaxial top layers, may lead to an overestimation of the layer roughnesses. In the diffraction experiments, the samples were measured with a combination of coupled $2\theta$-$\theta$ and rocking curve scans. Texture analysis was performed employing rotational $\phi$ scans at different sample tilts $\chi$. A pole figure was measured for $\phi$ angles between 350 and 190~degrees. Data in the range between 190 and 350~degrees in $\phi$ was assumed to be rotational symmetric with an angle of 180~degrees. The lattice mismatch between film and substrate for certain epitaxial relationships was calculated based on a previously established approach by \citet{Wildes2001}. Peak positions in $2\theta$ were determined by fitting with a Gaussian function, while rocking curve peaks were fitted with a Lorentzian profile. All error bars for fits of scattering data are statistical and do not include systematic errors arising from alignment or absorption. XRD patterns including Laue oscillations were additionally fitted with \textsc{GenL} \cite{Ravensburg2023Fit} enabling the determination of the average number of coherently scattering planes contributing to the Laue oscillations $N_\text{L}$ and the average out-of-plane atomic distance of $\alpha$-W $d_\text{hkl}$ as well as of a potentially present strain profile and layer/interface roughnesses.

Both, the Hall coefficient and resistivity were measured with a 4-point probe setup in a van der Pauw geometry, with the contact pins placed in the corners of a 8$\times$8~mm$^2$ square placed concentrically on the sample surface. The film itself had lateral dimensions of 10$\times$9~mm$^2$. Electronic transport measurements of the resistivity and Hall coefficient were performed during warm-up from 10 to 320~K (resistivity) and 20 to 320~K (Hall coefficient) in a cryostat using a closed cycle He compressor. The temperature was controlled stepwise using a 37~W resistance heater and a Cernox\textsuperscript{\textregistered} temperature sensor connected to a LakeShore 340 temperature controller. All measurements were done in thermodynamic equilibrium, by waiting until the temperature and voltage readings stabilised sufficiently for the respective measurements. The resistivity was measured at remanence using a 4-point van der Pauw method including reversed polarity measurements \cite{vanderPauw1958}. A current of 0.001~A was applied by means of a Keithley 2400 SourceMeter. A Keithley 2182A NanoVoltMeter was used to measure the voltage.

To measure the Hall coefficient, magnetic fields of -0.475 and 0.475~T were applied using a GMW Model 5403 electromagnet and a Kepco BOP 20-50MG power supply. The magnetic field was measured using a Hall probe and a LakeShore 455 Gaussmeter. The same 4-point probe setup as for the resistivity measurements was used to measure the Hall coefficient. To determine the Hall coefficient, the current was applied along one diagonal of the sample, while the Hall voltage was measured perpendicular to it. The Hall voltage was measured along both diagonals of the near square shaped sample and at positive and negative magnetic fields to account for geometric effects in the pin placement and sample geometry. In each magnetic field and current orientation, the current direction was alternated in a delta-measurement to account for electromotive forces \cite{vazquez2001}, thermally induced through minuscule temperature gradients inside the sample. A HP 3488A Switch/Control unit was used to automatically change pin connections for resistivity and the Hall coefficient measurements. The error bars for the resistivity measurements represent a statistical standard deviation of 20 repeated measurements, while the error bars for the Hall coefficient depict the propagated error of the statistical standard deviation of 6 successive voltage measurements in each geometry and field.

Scanning transmission electron microscopy (STEM) measurements were performed to combine reciprocal and real space information from the same spatial location of the sample at high resolution. Selected samples were examined in cross-section geometry using Titan Themis 200 from FEI operated at 200~000~V. The cross-section lamellae of W/CoFe bilayered films were prepared perpendicular to the side of the samples using a focused ion beam (FIB) Zeiss FIB/SEM Crossbeam 550 with Ga Ion-Sculptor gun system. The final polishing was performed at 5000~V ion acceleration voltage with XeF$_2$ gas assistance.

\subsection{\label{sec:dft}Density functional theory calculations}\label{subsec3}

Density functional theory (DFT) \cite{Hohenberg} was employed at 0~K to explore the atomic and electronic structure of two interfaces, namely W~($110$) on Al$_{2}$O$_{3}$~($11\bar{2}0$) with W~[$1\bar{1}1$]$\parallel$Al$_{2}$O$_{3}$~[$0001$] as well as W~($001$) on MgO~($001$) with W~[$100$]$\parallel$MgO~[$110$]. The Vienna {\it ab~initio} simulation package was used. The projector augmented wave potentials were chosen for the basis set \cite{Kresse1993, Kresse1994, Kresse1999} and the generalized gradient approximation, as parameterized by Perdew, Burke, and Ernzerhof \cite{Perdew}, was used to describe the exchange-correlation effects. The Bl{\"o}chl correction was employed \cite{Blochl} for the interfaces and an integration over the Brillouin zone was performed with the Monkhorst-Pack approach \cite{Monkhorst} with a {\it k}-point mesh of 4×4×1 for W~($110$) on Al$_{2}$O$_{3}$~($11\bar{2}0$) (198 atoms) and 8x8x1 for W~($001$) on MgO~($001$) (112 atoms). The convergence tests regarding the slab (substrate) thickness and {\it k}-points were carried out previously \cite{Sigumonrong2011}. Default values for the energy cut-off were taken into account. Considering the size of the Al$_{2}$O$_{3}$/W configuration (198 atoms), larger lateral configurations were not probed due to limited computational resources, but similar slab thicknesses, i.e., the number of stacked layers, for both interfaces were considered so that the results should be comparable. Six layers of W were always stacked on each substrate. The orthorhombic description of corundum Al$_{2}$O$_{3}$ was used to construct the interfaces \cite{Sigumonrong2011}. Oxygen termination of the Al$_{2}$O$_{3}$~($11\bar{2}0$) surface was assumed as it is reported to be a more likely match of the actual substrate surface in the experiments \cite{Mayer1990_2, Mayer1992, Guo1992, Blonski1993}. Our intention was to compare equivalent bonds across the interfaces (W – O for both W/Al$_{2}$O$_{3}$ and W/MgO rather than W – Al for W/Al$_{2}$O$_{3}$ and W – O for W/MgO). Hence, the calculated values of work of separation indicate likely trends to occur experimentally. Six atomic layers of W were taken into account, whereby W atoms were placed at the top position of O atoms. The convergence criterion for the total energy was 0.01~meV and a cut-off energy was 500~eV. All interfaces were constrained to the calculated lattice parameters of bulk Al$_{2}$O$_{3}$ and MgO at 0~K, acting as substrates. To construct these interfaces, a vacuum layer was inserted perpendicularly to the interface with a thickness of 10~{\AA}. The bottom layer of each substrate was frozen to mimic the infinite bulk. The interfaces were characterized by a work of separation $W_S$ \cite{Sigumonrong2011, Lin2007}, calculated from the total energy change per unit area upon separation of the corresponding slabs. All counterparts were fully relaxed at 0~K. The electronic structure was characterized by evaluating electron density distributions.

\section{\label{sec:results}Results and discussion}\label{sec3}
\subsection{\label{sec:structuralAnalysis}Growth of epitaxial W thin films}\label{subsec4}

\begin{figure*}[hbt!] 
\centering
\includegraphics[width=\textwidth]{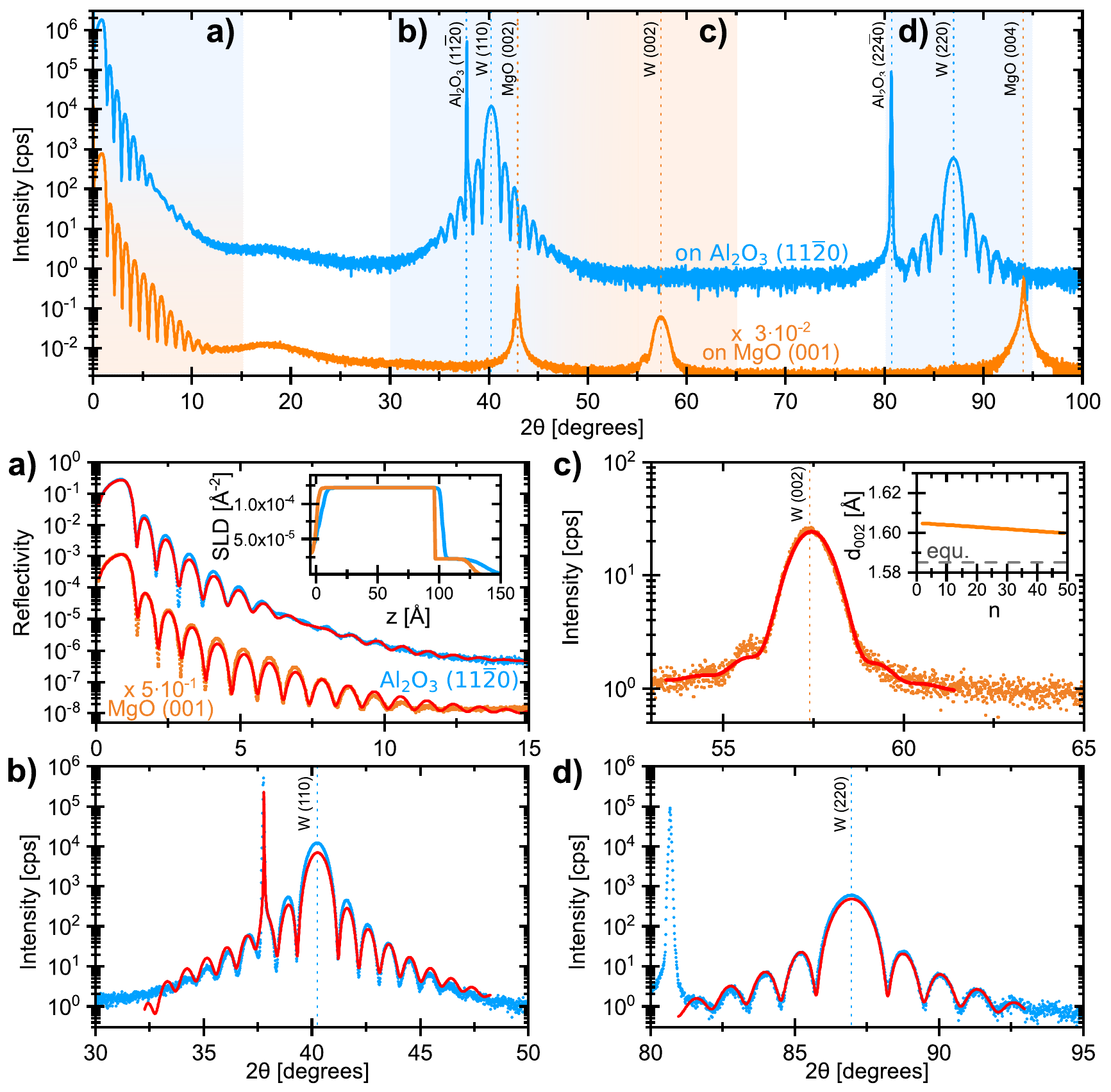}
\caption{X-ray scattering patterns of 100~{\AA} W layers grown on single crystalline Al$_{2}$O$_{3}$~($11\bar{2}0$) and MgO~($001$) substrates. The films are capped with 30~{\AA} Al. Indexed peaks relate to the $\alpha$-W structure. The highlighted regions in the upper panel are displayed as a) to d) in the panels below. Red curves correspond to fits of a) the reflectivity and b) to d) the indicated Bragg peaks indexed $hkl$. From the fits, a) the scattering length density $SLD$ profiles over thickness $z$ and c) the evolution of out-of-plane lattice spacing $d_{hkl}$ over $n$ lattice planes were obtained and are shown as insets. The grey dashed line corresponds to the $d_{hkl}$ spacing of the equilibrium $\alpha$-W structure \cite{Jain2013, osti_W}. For b) and d) no strain profile was applied. The scattering patterns are shifted vertically for clarity.}
\label{fig:fitting}\hfill
\end{figure*}

To compare the epitaxial growth of single layers of W on Al$_{2}$O$_{3}$~($11\bar{2}0$) and MgO~($001$) substrates, thin films, 100~{\AA} in thickness, were sputtered under the same deposition conditions. Their x-ray scattering patterns are displayed in the upper panel of Fig.~\ref{fig:fitting}. We start the discussion on the structural quality by having a closer look on the layering of the films, i.e., the low angle scattering regime displayed in detail in Fig.~\ref{fig:fitting}a. Both patterns show pronounced Kiessig fringes \cite{Kiessig1931} up until 10 and 12 degrees in 2$\theta$ for the films grown on Al$_{2}$O$_{3}$~($11\bar{2}0$) and MgO~($001$), respectively. The presence of Kiessig fringes up to 10 degrees is typically only found in layers that are flat on a mesoscopic length scale of the order of the in-plane coherence length of the x-ray beam. Scattering length density profiles obtained from the fitting of the reflectivity are shown as insets and confirm the intended substrate/W/Al layering with well defined layer thicknesses on both substrates.

For the thin film grown on Al$_{2}$O$_{3}$~($11\bar{2}0$), the interface widths and surface roughness of the substrate, the 97(1)~{\AA} W layer, and the 35(1)~{\AA} Al layer are 2(1), 2(1) and 9(1)~{\AA}, respectively. The interface widths and surface roughness of the MgO~($001$) substrate, the 97(1)~{\AA} W layer, and the 30(1)~{\AA} Al layer are 2(1), 0(1), and 4(1)~{\AA}, respectively. Furthermore, for Al$_{2}$O$_{3}$~($11\bar{2}0$), the reflectivity data could only be fitted by including an additional 5(1)~{\AA} layer with a roughness of 3(1)~{\AA} at the substrate/W interface in the fit, with electronic density close to tungsten oxide. Since roughness and thickness of this layer are of the same order of magnitude, the roughness value may not be indicative of the real underlying roughness. A similar observation has been made for Nb growing on sapphire, relating the presence of an oxide layer to a kinematic chemical reaction between a film and a substrate \cite{Wildes2001}. The observation is in line with the reported bonds forming between W atoms growing on Al$_{2}$O$_{3}$ and the oxygen atoms of the substrate \cite{Park2022}. As the nominal thicknesses for all samples lie reasonably close ($<$~4~\% for W and CoFe) to the fitted layer thicknesses, we will continue to refer to the nominal thickness henceforth.

Regarding the diffraction analysis of the W thin film grown on Al$_{2}$O$_{3}$~($11\bar{2}0$), two peaks corresponding to the ($110$) and ($220$) $\alpha$-W structure Bragg reflections are visible at 40.254(0) and 86.962(1)~degrees, corresponding to the out-of-plane atomic distances $d_{110}$ and $d_{220}$ in W of 2.239(0) and 1.119(0)~{\AA}, respectively. Since the Bragg peaks of the $\beta$-W structure lie within a few degrees of the observed ($110$) and ($220$) peaks \cite{Petroff1973, OKeefe1996, Liu2007}, reciprocal space mapping was conducted around the $\alpha$-W~($002$) reflection at 58~degrees \cite{Chen2022, Jain2013, osti_W} in 2$\theta$. At a tilt $\chi$ of the sample by 45~degrees perpendicular to the scattering plane, we observed a sharp peak, confirming the phase-pure epitaxial growth of $\alpha$-W on Al$_{2}$O$_{3}$~($11\bar{2}0$) with [$110$] out-of-plane growth direction under the above mentioned conditions. For simplicity, the $\alpha$-W is referred to as W for growth on Al$_{2}$O$_{3}$~($11\bar{2}0$) henceforth. The sharp peaks, namely Al$_{2}$O$_{3}$~($11\bar{2}0$) and ($22\bar{4}0$), are attributed to the substrate, while the broad bump at around 18~degrees is attributed to the amorphous capping layer.

In addition, around the W~($110$) and ($220$) Bragg peaks, magnified in Fig.~\ref{fig:fitting}b and d, respectively, symmetric Laue oscillations are visible over a range of more than 10~degrees. The occurrence of the Laue oscillations is proof of a high degree of coherent scattering and, therefore, high crystal quality over the total thickness of the W layer \cite{Lee2017, Wildes2001}. As defects and dislocations give rise to coherent diffuse scattering and do not contribute to the observed intensity of these oscillations, the shape and decay of the Laue oscillations can be used as a quantitative measure for the crystal quality of epitaxial thin films \cite{Lee2017}. The diffracted intensity around the W~($110$) and ($220$) Bragg peaks was fitted in order to identify the degree of coherent scattering. Fits are shown as red lines in the respective figures. The symmetry of the oscillations indicates a negligible degree of strain in the 100~{\AA} thick W layer \cite{Ravensburg2022}, which is in line with the reports on epitaxial W growth on sapphire by pulsed laser deposition \cite{Grath1994}. Hence, no strain profile was included in the fitting for the sample grown on Al$_{2}$O$_{3}$~($11\bar{2}0$).

As it is evident from Fig.~\ref{fig:fitting}, the fitting captures the features of the diffraction pattern. The main ($110$) Bragg peak intensity however is not entirely captured, with the fitted interface roughness being underestimated. Based on both fits, 99.9~\% of the pure W layer scatters coherently. It has to be noted, however, that the thickness of the tungsten oxide resembling interface layer in the XRR fitting was not included in this calculation due to its unknown crystal structure. Including it by assuming the obtained average W layer spacing of the film above, yields a reduced percentage of 94.7~\%. The fitted average out-of-plane distances $d_{110}$ and $d_{220}$ lie within 0.1~\% of the previously determined values. The interface roughnesses, contributing to the decay of the intensity of the Laue oscillations with increasing angular distance from the W~($110$) and ($220$) main peaks, are fitted to be 2 and 5~{\AA}, respectively, and thus in agreement with the fitted roughnesses based on the reflectivity data.

On MgO~($001$), the only observed specular W peak is at 57.404(2)~degrees and corresponds to a $d_{002}$ of 1.604(0)~{\AA} of the $\alpha$-W structure. For simplicity, $\alpha$-W is henceforth also referred to as W for growth on MgO~($001$). The sharp MgO~($002$) and ($004$) peaks are attributed to the single crystalline substrate. W is growing epitaxially in the [$001$] growth direction on MgO~($001$), in line with results from previous studies \cite{Zheng2015}. However, in this study, Laue oscillations are observed on the low angle side, observable in the magnified display of the W~($002$) Bragg peak in Fig.~\ref{fig:fitting}c. The oscillations are less pronounced, decay faster, and are more asymmetric compared to W grown on Al$_{2}$O$_{3}$~($11\bar{2}0$). In the fitting, the asymmetry is accounted for by a strain profile. The fitted variation of the out-of-plane spacing $d_{002}$ as a function of lattice planes $n$ across the W layer thickness is shown in the inset. The out-of-plane spacing seems to be linearly decreasing over W layer thickness, lying around 2~\% above the equilibrium lattice spacing of bulk W \cite{Jain2013, osti_W}. Lattice mismatch between film and substrate gives rise to misfit strain in epitaxial thin films causing a change in out-of-plane lattice spacing over film thickness. The origin of strain in W grown on MgO~($001$) will be discussed below. Based on the fitting, 80.4~\% of the W layer grown on MgO~($001$) scatter coherently. The fitted $d_{002}$ lies within 0.1~\% from the previously determined value, and the interface roughness of 5~{\AA} is in agreement with the roughness obtained from XRR.

\begin{figure}[t!] 
\centering
\includegraphics[width=\columnwidth]{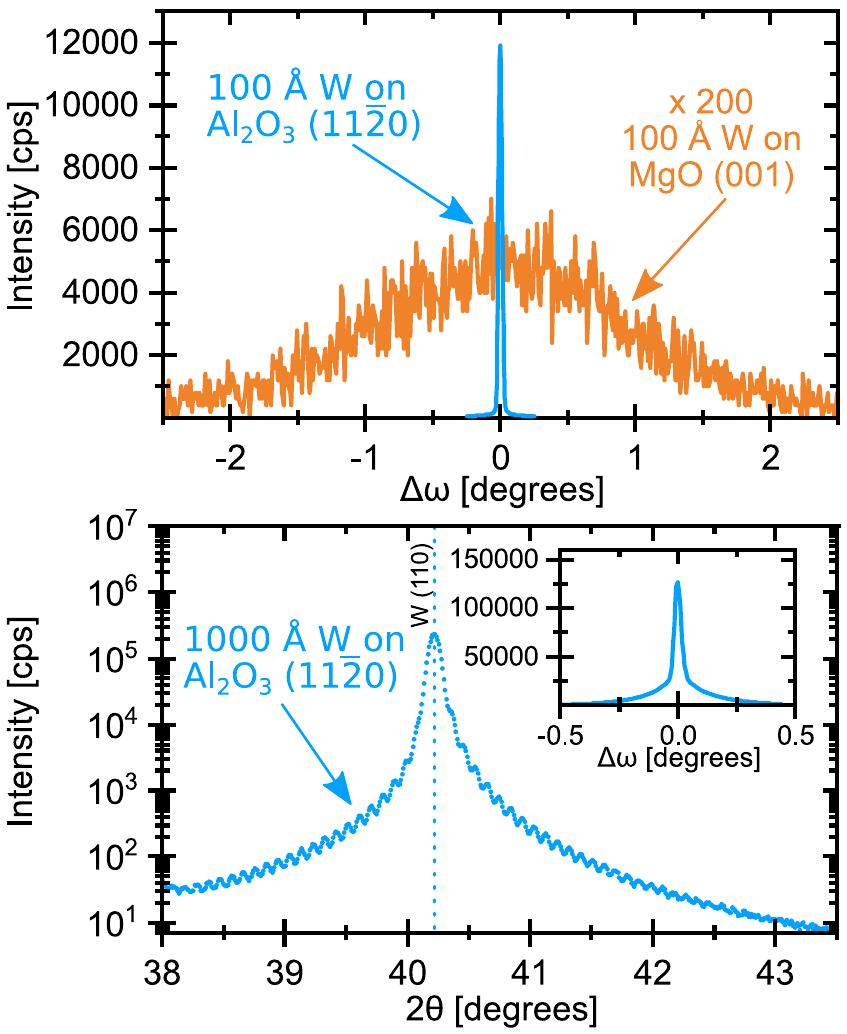}
\caption{Top panel: Rocking curves measurements around the specular $\alpha$-W Bragg peak of 100~{\AA} W layers grown on single crystalline Al$_{2}$O$_{3}$~($11\bar{2}0$) and MgO~($001$) substrates. For Al$_{2}$O$_{3}$ the measurement was conducted around the W~($110$) Bragg peak and for MgO the measurement was conducted around the W~($002$) Bragg peak. In order to visually compare the widths, the rocking curve of the sample grown on MgO was multiplied by a factor of 200. Bottom panel: X-ray diffraction pattern around the W~($110$) Bragg peak of 1000~{\AA} W grown on single crystalline Al$_{2}$O$_{3}$~($11\bar{2}0$). The rocking curve for this peak is displayed in the inset.} 
\label{fig:thickness}\hfill
\end{figure}

The degree of coherent scattering of the W layer can, as discussed above, be used as a quantitative measure of the crystal quality of epitaxial W, which is higher for Al$_{2}$O$_{3}$~($11\bar{2}0$) than that for MgO~($001$). Moreover, the observed difference in the crystal quality manifests itself in the peak intensity distribution in reciprocal space. The peak intensity of the Bragg peaks is one to three orders of magnitude higher for W on Al$_{2}$O$_{3}$~($11\bar{2}0$) compared to W on MgO~($001$). This difference is related to a larger mosaic spread of W grown on MgO~($001$), as it can be seen in the upper panel of Fig.~\ref{fig:thickness}. The rocking curve around the W~($110$) peak for W grown on Al$_{2}$O$_{3}$~($11\bar{2}0$) is sharp and has a full width at half maximum (FWHM) of 0.02(0)~degrees. The peak intensity of the rocking curve around the W~($002$) peak for W grown MgO~($001$) is distributed over a two orders of magnitude wider angular range (FWHM = 2.19(5)~degrees). The instrument resolution of 0.012~degrees was taken into account for determining these values. The mosaic spread, i.e. misorientation of the W atomic planes relative to each other, is larger, corresponding to a lower crystal quality of the film grown on MgO~($001$).

As the mosaic spread of the epitaxial W thin film on Al$_{2}$O$_{3}$~($11\bar{2}0$) is low, it is possible to grow relatively thick films still exhibiting a high degree of coherent scattering. The diffractogram around the W~($110$) Bragg peak of a 1000~{\AA} thick W layer is shown in the lower panel of Fig.~\ref{fig:thickness}. Even for this tenfold larger layer thickness, W grows epitaxially with [$110$] growth direction on Al$_{2}$O$_{3}$~($11\bar{2}0$). The Bragg peak position relates to an average $d_{110}$ spacing over 1000~{\AA} of 2.240(0)~{\AA}, a deviation of less than 0.08~\% from $d_{110}$ for 100~{\AA} W. The Laue oscillations observed on both sides are symmetric and can be observed over a range of more than 4~degrees in 2$\theta$, being proof of a small degree of strain over the W layer thickness. To the best of the authors' knowledge, the Laue oscillations of W have only been observed for thin films of 30~{\AA} layer thickness \cite{Grath1994}. The existence of the Laue oscillations for a film thickness of 1000~{\AA} is proof of the superior crystal quality of the W films deposited in this study and supports the argument for the fast relaxation at the interface once again. It showcases that the crystal quality is not deteriorating with thickness and up until 1000~{\AA} thickness there is no onset of relaxation through dislocations. The FWHM of 0.04(0)~degrees of the ($110$) rocking curve for 1000~{\AA} W, shown in the inset, is comparable to the value for 100~{\AA} W. However, the shape of the rocking curve for 1000~{\AA} W includes two features indicating two different correlation lengths; a narrow feature, almost resolution limited, and a broader triangularly shaped feature. Similar observations of features in rocking curves have been reported for epitaxial Nb~($110$) grown on Al$_{2}$O$_{3}$~($11\bar{2}0$) \cite{Wildes2001}, where the broad feature is more pronounced for thicker films. For Nb, \citet{Wildes2001} show that flat growth planes over long length scales give rise to the narrow component, while strain and misfit dislocations causing height deviations give rise to the broader feature in the rocking curve. The integrated intensity of the broad component is thickness dependent for Nb, which is attributed to an increased layer roughness for thicker films \cite{Woelfing1999}. The broad component can be viewed as related to the defect density. For Nb, a large part of the relaxation takes place at the substrate/Nb interface, i.e. the growth is semicoherent. Due to the similarity between the observed scattering from Nb($110$) and W($110$) thin films grown on Al$_{2}$O$_{3}$~($11\bar{2}0$), semicoherent growth is likely to also be present in W thin films.

\begin{figure}[b!] 
\centering
\includegraphics[width=\columnwidth]{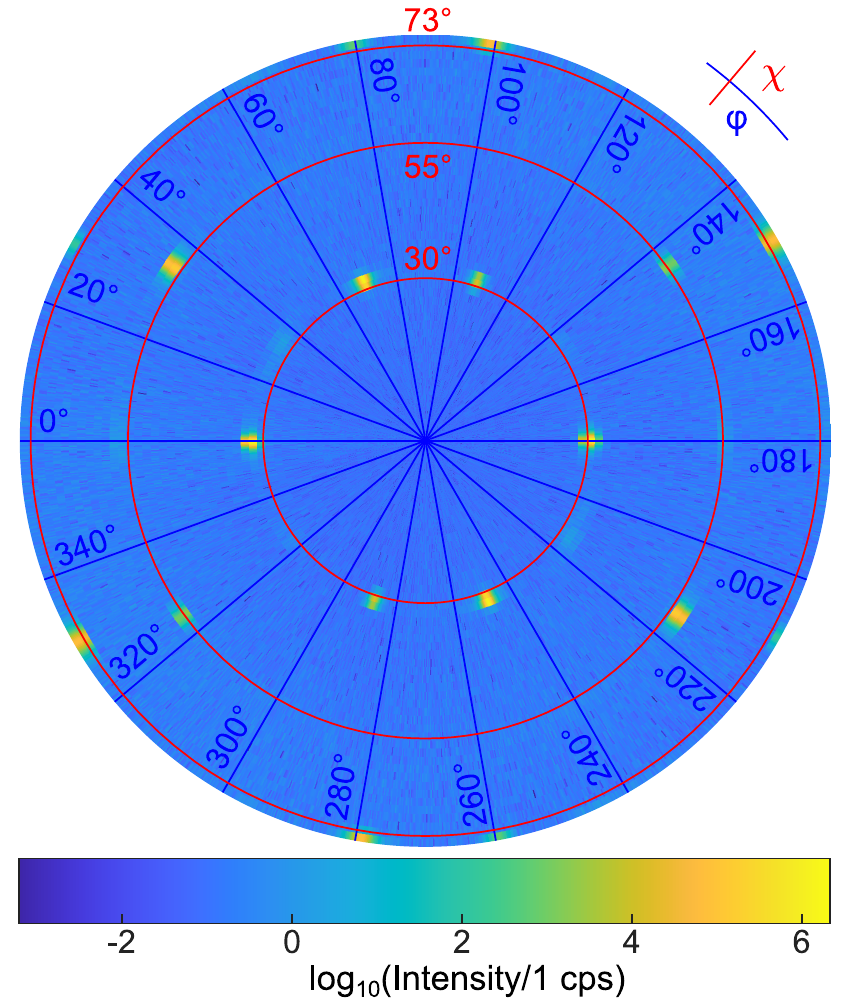}
\caption{W~$\{112\}$ x-ray pole figure measured on a 1000~{\AA} W layer grown on single crystalline Al$_{2}$O$_{3}$~($11\bar{2}0$). The pole figure displays the polar angle $\chi$ (sample tilt) and the azimuthal angle $\phi$ (sample rotation). Specific $\chi$ and $\phi$ values relevant for the discussion are marked as red circles and blue lines, respectively.} 
\label{fig:polefigure}\hfill
\end{figure}

Hetero-epitaxy is restricted to specific relative orientations of substrate and film, as lattice matching is required. On Al$_{2}$O$_{3}$~($11\bar{2}0$), epitaxial W~($110$) growth is observed due to a match between these two crystal planes at the substrate interface. The W~($110$) crystal plane has a rectangular atomic shape with an atomic distance $d_{110}$ on one side and $d_{001}$ on the other side. The Al$_{2}$O$_{3}$~[$0001$] and [$1\bar{1}00$] directions span the corresponding Al$_{2}$O$_{3}$~($11\bar{2}0$) plane \cite{Souk1987,Grath1994}. It is reported that a W~($110$) growth orientation with an in-plane rotated unit cell is energetically favored to match a rectangular structure on the Al$_{2}$O$_{3}$~($11\bar{2}0$) surface \cite{Souk1987,Grath1994, Wildes2001, Mayer1990}.

Based on the work of \citet{Grath1994}, this rotation is calculated to be 54.7~degrees relative to Al$_{2}$O$_{3}$~[$0001$] or 35.3~degrees relative to [$1\bar{1}00$]. To further investigate the in-plane orientation of the W unit cell in our thin films, a W~$\{112\}$ pole figure measurement was conducted on a 1000~{\AA} thick W layer. The results are displayed in Fig.~\ref{fig:polefigure}. The off-specular W~$\{112\}$ peaks are expected to be observed in diffraction at the incident angle $\theta$ corresponding to $d_{112}$. However, to obtain a W~$\{112\}$ plane in diffraction, the sample needs to be tilted by a certain polar angle $\chi$ and rotated by the azimuthal angle $\phi$ based on the in-plane orientation of the unit cell relative to ($110$) out-of-plane orientation. The pole figure displays the polar angle $\chi$ (sample tilt) and the azimuthal angle $\phi$ (sample rotation). At $\phi$ = 0~degrees, the sample edges, corresponding to the Al$_{2}$O$_{3}$~[$0001$] and [$1\bar{1}00$] directions, are oriented 90 and 180~degrees to the incoming x-ray beam. Sharp peaks are observed at specific $\chi$ and $\phi$ angles, confirming epitaxial growth of 1000~{\AA} W on Al$_{2}$O$_{3}$~($11\bar{2}0$). The angles between [$110$] and $<$112$>$ are all either 30.0, 54.7, 73.2, or 90.0~degrees, depending on the specific crystallographic plane from the $<$112$>$ family. These $\chi$ angles are displayed as red circles. Within the resolution of the measurements related to the angles $\chi$ and $\phi$, the $\{112\}$ peaks are observed at these specific sample tilts, in line with the expected small degree of strain in the epitaxial W film. Moreover, a $\{112\}$ peak is observed at $\phi$ = 0 and 180~degrees for $\chi$ = 30.0~degrees. Therefore, a W~$<$112$>$ crystallographic direction is assumed to be parallel to the edge of the substrate, which is either [$0001$] or [$1\bar{1}00$] spanning the Al$_{2}$O$_{3}$~($11\bar{2}0$) plane.

In order to determine the relative crystal orientation of the substrate and the film, an Al$_{2}$O$_{3}$~$\{11\bar{2}0\}$ pole figure was measured for $\phi$ = -10 to 110~degrees and $\chi$ = 0 to 80~degrees. The only visible $\{11\bar{2}0\}$ peak measured within this range is at $\phi$ = 0~degrees and for a sample tilt of $\chi$ = 60~degrees. In a hexagonal crystal with ($2\bar{1}\bar{1}0$) orientation, ($11\bar{2}0$) satisfies the diffraction criterion for a 60~degrees sample tilt $\chi$ with [$0001$] rotation axis. Therefore, at $\phi$ = 0~degrees, [$0001$] is parallel to a W~$\{112\}$ plane and thus perpendicular to the respective $<$112$>$ direction. Hence, our experimental diffraction study confirms the epitaxial relationships reported by \citet{Grath1994}: W[$1\bar{1}1$]$\parallel$Al$_{2}$O$_{3}$[$0001$] and W[$\bar{1}12$]$\parallel$Al$_{2}$O$_{3}$[$1\bar{1}00$] or W[$1\bar{1}\bar{1}$]$\parallel$Al$_{2}$O$_{3}$[$0001$] and W[$1\bar{1}2$]$\parallel$Al$_{2}$O$_{3}$[$1\bar{1}00$].

\begin{figure}[t!] 
\centering
\includegraphics[width=\columnwidth]{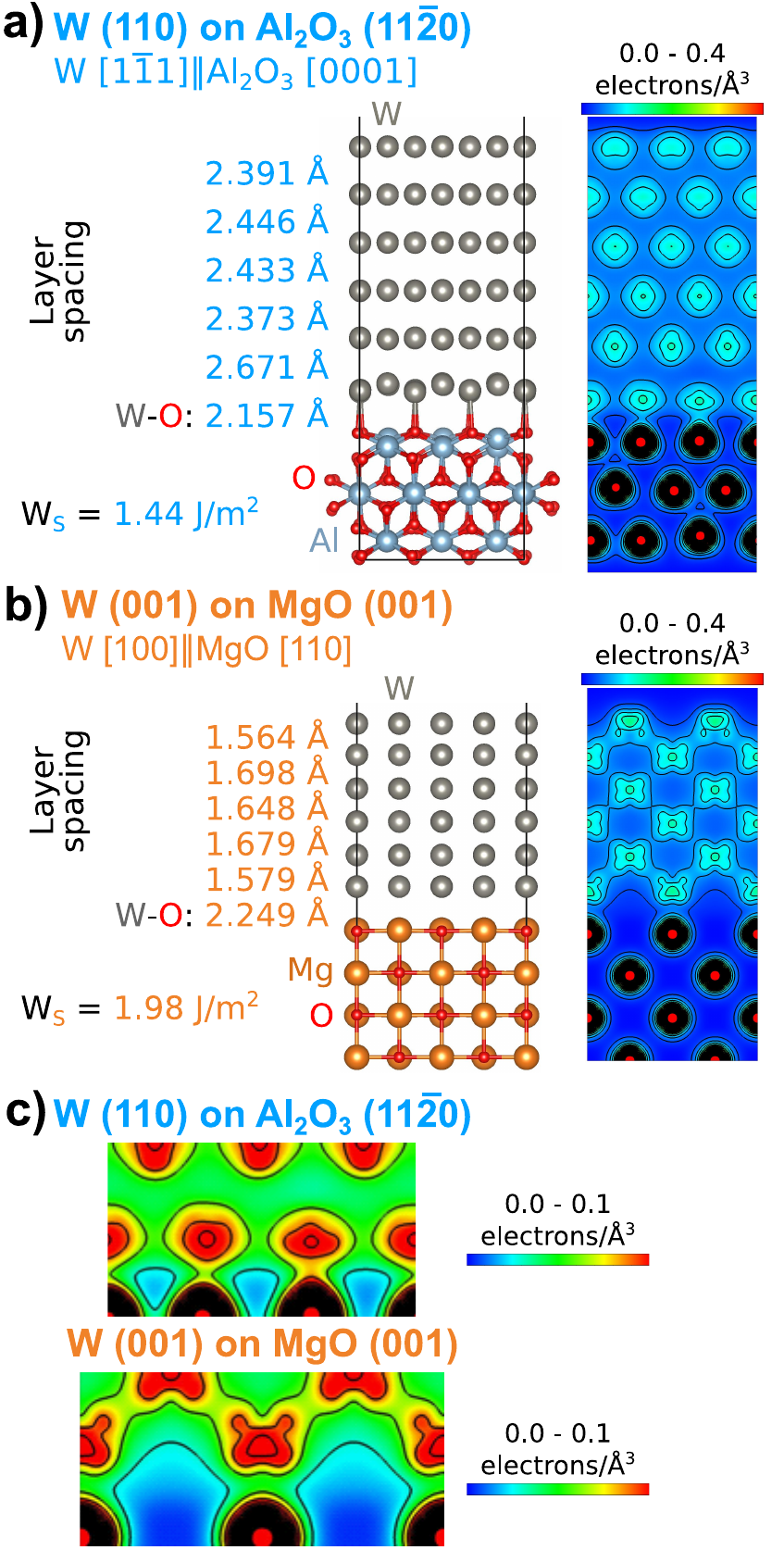}
\caption{Results of density functional theory calculations of 6 monolayers of a) W~($110$) on Al$_{2}$O$_{3}$~($11\bar{2}0$) with W~[$1\bar{1}1$]$\parallel$Al$_{2}$O$_{3}$~[$0001$] and b) W~($001$) on MgO~($001$) with W~[$100$]$\parallel$MgO~[$110$]. A schematic of the atomic structure and the electron density distributions of the interface are displayed in the middle and to the right of each figure, respectively. In-plane and out-of-plane atomic distances at the interfaces and in the W layers are indicated. The work of separation $W_S$ is displayed on the left side. For Al$_{2}$O$_{3}$, oxygen surface termination is assumed. c) Zoom-in of the electron density map onto the substrate/W interface.}
\label{fig:dft}\hfill
\end{figure}

The lattice mismatch for these epitaxial relationships are 7.2~\% and 19.4~\% along W~[$1\bar{1}1$] and W~[$\bar{1}12$] directions, respectively \cite{Jain2013, osti_W, osti_Al2O3}. The mismatch is expected to cause misfit strain in the growing W layer, eventually leading  to the formation of misfit dislocations and strain release above the critical thickness for fully coherent growth. Due to the positive Poisson ratio of 0.284, determined from reported elastic constants \cite{Renault1998}, the strain in W is expected to be tensile in-plane and compressive out-of-plane since both atomic distances in Al$_{2}$O$_{3}$ are larger compared to the corresponding atomic distances in bulk W. For comparison, for the  epitaxial growth of similar sized Nb~($110$) on Al$_{2}$O$_{3}$~($11\bar{2}0$), lattice mismatches of 1.9~\% and 12.9~\% are reported in the two different crystallographic directions \cite{Mayer1990} and the critical thickness is reported to be less than 100~{\AA} \cite{Wildes2001}. Hence, for the W films grown within this study, we expect the formation of misfit dislocations for strain release giving rise to coherent diffuse scattering, which reduces the coherence length of the epitaxial crystal \cite{Ying2009}. However, parts of this dislocation formation is expected directly at the substrate/film interface for two reasons: First, in the stated epitaxial relationship, some W atomic positions do not coincide with atomic positions of the Al$_{2}$O$_{3}$ lattice, but lie close to octahedral interstices in the Al$_{2}$O$_{3}$ lattice \cite{Grath1994, Wildes2001}. Therefore, additional atomic relaxation is expected at the interface. Second, a miscut is common for the Al$_{2}$O$_{3}$~($11\bar{2}0$) substrates \cite{Yoshimoto1995, Wildes2001} that causes atomic steps and terraces with an incommensurate step height of $d_{0006}$. Therefore, the terraces will propagate into the growing film as defects. In between the defects, coherent regions with well-defined translational order are expected \cite{Wildes2001}. The presence of defects at the substrate/film interface is supported by the necessity to include an additional layer into the XRR fitting for both W and Nb \cite{Wildes2001} grown with ($110$) texture on Al$_{2}$O$_{3}$~($11\bar{2}0$). The defects as well as the additional XRD peaks are expected to reduce the misfit strain in W already at the interface, below the critical thickness. For Nb~($110$) on Al$_{2}$O$_{3}$, residual epitaxial strain which depends on the layer thickness is reported to lie usually below 0.05~\% \cite{Song1997}. Hence, growth with reduced misfit strain above the interface is expected for W on Al$_{2}$O$_{3}$~($11\bar{2}0$), in line with the result that 99.9~\% of the relaxed W layer on top scatter coherently.

The conundrum of obtaining epitaxy despite a large misfit strain requires further investigation. For that reason, DFT calculations of the epitaxial Al$_{2}$O$_{3}$~($11\bar{2}0$)/W~($110$) interface structure were performed. The results on the crystal structure including the corresponding electron density distribution are displayed in Fig.~\ref{fig:dft}a. Based on these, strain is introduced into the growing W layer at the interface, as evident from the tetragonal distortion of the W slab. To match the in-plane atomic distances in the substrate, the in-plane atomic spacing $d_{1\bar{1}0}$ at the interface was calculated to be 2.774~{\AA}, an increase of 23~\% compared to 2.242~{\AA} \cite{Jain2013, osti_W} at equilibrium, which is in line with the expected in-plane tensile strain. In the out-of-plane direction, the calculations predict a significantly larger $d_{110}$ = 2.671~{\AA} between the first and second atomic layers compared to the interplanar spacing of the following monolayers. Furthermore, the DFT calculations indicate that the first atomic layer of W exhibits a buckled atomic structure. The center atom in the ($110$) atomic plane has a larger distance to the interface as compared to the oxygen bound corner atoms in the unit cell. Such buckling often occurs due to high interfacial strains \cite{Cheng_APL2016}. The exceptionally large interplanar spacing between the first and second atomic W layer in combination with the buckling of the atomic structure can be assigned to different chemical and structural properties of W directly at the interface, in agreement with the tungsten oxide resembling interface layer included in the XRR fitting as well as the expected strain relaxation at the interface. From the electron density distribution of the interface structure, a change in electron density of the first three monolayers is evident, showing an out-of-plane elongation of the area of high electron density around the atomic positions. First-principles calculations on the similar Al$_{2}$O$_{3}$~($0001$)/Nb~($111$) system also show appreciable interlayer relaxation near the interface for an oxygen terminated substrate surface \cite{Kruse1996}. However, all calculated atomic distances in W are larger than the equilibrium $d_{011}$ of 2.242~{\AA} \cite{Jain2013, osti_W}. This out-of-plane elongation is in contrast to the expected compressive strain in this direction and to the experimentally observed smaller $d_{011}$ in the out-of-plane direction of around 2.239(0) and 2.240(0)~{\AA} for the 100 and 1000~{\AA} thick W layers, respectively. It should be remarked that the high interfacial strains and hence high atomic relaxations are partly due to the interface size. Classical molecular dynamics modelling may reveal these particularities, but we are not aware of any available interatomic potentials. 

The calculated work of separation $W_S$ for the  Al$_{2}$O$_{3}$/W~($110$) system with W~[$1\bar{1}1$]$\parallel$Al$_{2}$O$_{3}$[$0001$] is 1.44~J/m$^2$, corresponding  to an intermediately strong interface where epitaxy may be possible. For substrate/film combinations known for their epitaxial growth like Nb($111$) on Al$_{2}$O$_{3}$($0001$) or Cu($111$) on Al$_{2}$O$_{3}$($0001$) larger values for $W_S$ are reported, namely 12.7 \cite{Batirev1999} and 5.48~J/m$^2$ \cite{Herrig2018}, respectively. A comparable work of separation of 2.86~J/m$^2$ is reported for V$_2$AlC($0001$) on Al$_{2}$O$_{3}$~($11\bar{2}0$) \cite{Sigumonrong2011}. The interface of Al$_{2}$O$_{3}$~($11\bar{2}0$)/V$_2$AlC($0001$) was characterized as semicoherent, i.e. having coherent regions separated by misfit dislocations due to a lattice mismatch of 8.16~\% \cite{Sigumonrong2011}.

\begin{figure}[t!] 
\centering
\includegraphics[width=\columnwidth]{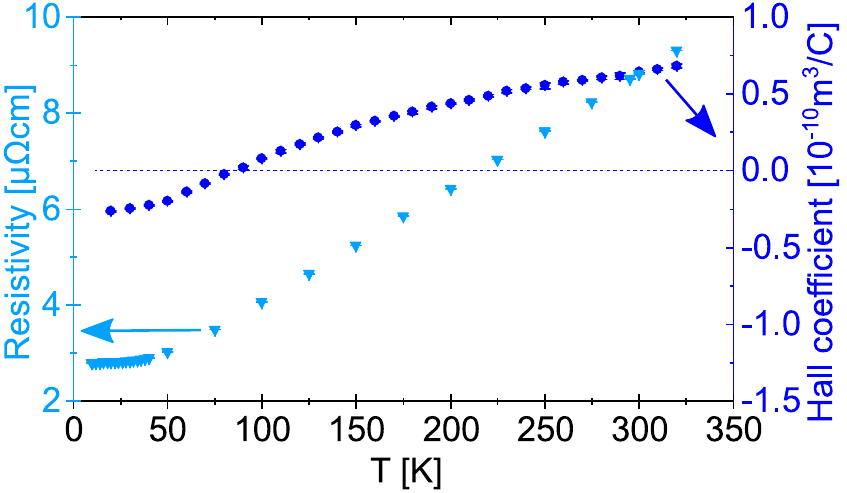}
\caption{Temperature dependent electronic resistivity and Hall coefficient measurements of a 1000~{\AA} thick W layer grown on Al$_{2}$O$_{3}$~($11\bar{2}0$).}
\label{fig:transport}\hfill
\end{figure}

Results of temperature dependent electronic resistivity measurements of 1000~{\AA} W grown on Al$_{2}$O$_{3}$~($11\bar{2}0$) are displayed in Fig.~\ref{fig:transport}. The statistical error of each data point is of the order of $1\times10^{-4}$~$\mu \Omega\textrm{cm} $ and thus smaller than the symbols depicted. The data exhibits a typical Bloch-Gr{\"u}neisen metallic behaviour with a residual resistivity ratio (RRR) of 3.14. The ratio is comparable to the value reported by \citet{Choi2012}, with a RRR of around 4 for annealed samples. Hence we conclude that the defect density in the present samples are comparable to the state of the art of epitaxial tungsten films grown with sputtering. While the previous discussion of the XRD data strongly suggests excellent crystal quality, the measured RRR in this case remains relatively low. One possible explanation for this disparity could lie in the origin of the XRD and resistivity signals. The XRD intensities arise from coherent scattering within the coherence volume of the x-ray beam, typically limited to a few thousands {\AA}ngstr{\"o}ms. Conversely, the resistivity signal probes the entire sample area, with the probing pins separated by several millimeters.

The Hall coefficient in the present sample is about half of the bulk value of 11.3$\times 10^{-11}$~m\textsuperscript{3}/C \cite{Frank1957} at 300~K and decreases with decreasing temperature. In contrast to the bulk sample of \cite{Frank1957}, it exhibits a crossover close to 85~K. Care is needed when comparing results of measurements on single crystals to polycrystalline materials, since the Hall effect depends on the crystallographic direction \cite{Hurd1972, Frank1957, Volkensthein1964}. An increase in the crystallographic defect density is expected to increase the resistivity and alter the Hall coefficient. However, the influence of defects on the Hall coefficient are due to their impact on the anisotropy of the scattering rates rather than their absolute value \cite{schulz1992}. Furthermore, the Fuchs-Sondheimer model of surface scattering predicts an increase of resistivity and the Hall coefficent with decreasing film thickness \cite{Choi2012, Panchenko1969} with the present film being reasonably close to the bulk regime, where finite size is not seriously affecting the results. The behaviour of the Hall coefficient versus temperature in Fig.~\ref{fig:transport} exhibits the opposite trend as that of $\alpha$-W films grown on thermally oxidized Si, as reported by \citet{betaW_APL}, whereas $\beta$-W, exhibits consistently negative Hall coefficients at temperatures between 10 and 300~K \cite{betaW_APL}. This interesting difference in the sign and the magnitude of the Hall effect as compared to \cite{betaW_APL} might be due to the large difference in defect densities, phase purity, and/or crystallographic orientation, between the epitaxially grown films here and the textured films grown by \citet{betaW_APL}. In fact, \citet{bastl1972} concluded from Hall measurements on thin polycrystalline films of W that the presence of grain boundaries together with defects resulted in a suppression of the Hall coefficient as compared to the bulk value. The sign of the Hall coefficient in the present work was confirmed (at ambient temperature) by comparing the results of measurements of the Hall coefficient in the same setup of polycrystalline bulk samples to the literature.

In contrast to the epitaxial W thin film grown on Al$_{2}$O$_{3}$~($11\bar{2}0$), W grown on MgO~($001$) exhibits the [$001$] out-of-plane growth direction. Epitaxial growth of W on MgO~($001$) was confirmed as off-specular peaks were measured as sharp reflections occurring at specific sample tilts $\chi$ and rotations $\phi$ in reciprocal space. Measurements of the off-specular W~$\{112\}$ reflections revealed a relative in-plane rotation of 45~degrees between the W~[$100$] and MgO~[$100$] directions, i.e. W~[$100$] is oriented parallel to MgO~[$110$], confirming the observed epitaxial relationship reported elsewhere \cite{Zheng2015}. This in-plane rotation allows for a smaller mismatch of around -6.1~\% between the respective atomic distances in film and substrate \cite{Jain2013, osti_W, Zheng2015}. At elevated deposition temperatures like 843~K, the lattice mismatch is expected to be even smaller due to different thermal expansion coefficients of substrate and film \cite{Zheng2015}. Such a lattice rotation has been reported for other epitaxially growing thin films on MgO~($001$) substrates with similar atomic distances as W, e.g. Fe \cite{Vassent1996, Meyerheim2001, Meyerheim2002}, V, Cr, Hf \cite{Schroeder2015} or alloys thereof \cite{Devolder2013}. The mismatch is negative \cite{Wildes2001}, yielding compressive in-plane and tensile out-of-plane strain in W and hence, a tetragonally distorted unit cell. It is the opposite strain state as compared to W grown on Al$_{2}$O$_{3}$~($11\bar{2}0$). Hints of the expected tetragonal distortion of the unit cell can be found in the asymmetry of the Laue oscillations for W grown on MgO~($001$), displayed in Fig.~\ref{fig:fitting}c. The critical thickness for the introduction of misfit dislocations of W grown at 1173~K on MgO~($001$) with a mismatch of -5.1~\% is reported to be 27~{\AA}, well below the W thicknesses in this study. Hence, misfit dislocation formation is expected for all samples of W grown on MgO~($001$). The variation of the out-of-plane atomic distance $d_{200}$ over the thickness of the W layer lies above the equilibrium spacing for relaxed W \cite{Jain2013, osti_W}, in line with tensile out-of-plane strain. Over the layer thickness, $d_{002}$ is decreasing towards the equilibrium value, indicating a partial strain relaxation. 

These observations are in agreement with DFT calculations on the crystal structure of the MgO~($001$)/W~($001$) interface. The results of these calculations are displayed in Fig.~\ref{fig:dft}b. The in-plane $d_{100}$ lattice distance is calculated to be 3.004~{\AA} and, thus, smaller than the equilibrium value \cite{Wheeler1925,Jain2013, osti_W}, but in line with the expected compressive in-plane strain. It appears that W is tetragonally distorted for the first six monolayers, whereby the out-of-plane interplanar spacing is smaller than that of the equilibrium configuration, as expected for the tensile strain state in this direction. The spacing oscillates between roughly 1.60 and 1.69~{\AA} for every other layer. This highlights that the expected strain might affect atoms at different positions of the bcc cell differently, possibly due to different in-plane positions in relation to the substrate. Another reason may be the employed registry, since lateral relaxations may partly be inhibited due to the size of interface considered by DFT.

The calculated work of separation $W_S$ is similar to the value calculated for W~($110$) on Al$_{2}$O$_{3}$~($11\bar{2}0$). The interface is thereby characterized to be intermediately strong, allowing for epitaxy, possibly locally or up to low W thicknesses. The interfacial bonds between tungsten and the oxygen atoms of the substrate, however, are weaker in the case of MgO~($001$). This difference is also visible in the electron density map. In the case of both interfaces, the interfacial bonds are characterized by covalent contributions (charge sharing) and ionic contributions (charge transfer). Fig.\ref{fig:dft}c contains a zoom-in of the electron density distribution at the substrate/W interfaces. Up to 30~\% more charge is shared between W and O across the W/Al$_{2}$O$_{3}$ interface compared to W and O across the W/MgO interface implying a stronger covalent interaction. The bonds between W and O atoms across the Al$_{2}$O$_{3}$/W interface are shorter and thus stronger than those in the case of MgO, but a more transparent comparison should be made by comparing the bond lengths in the substrates (bulk counterparts). The Mg-O bond length is 2.124~{\AA}, while the corresponding interfacial bond (W-O) is 2.249~{\AA}. On the other hand, the Al-O bond length in the substrate is 1.874~{\AA} and the corresponding W-O bond is 2.157~{\AA}. Hence, a stacking sequence together with an expected bond length is better reproduced for W on MgO~($001$), which is mirrored in the slightly higher work of separation.

\begin{figure}[t!] 
\centering
\includegraphics[width=\columnwidth]{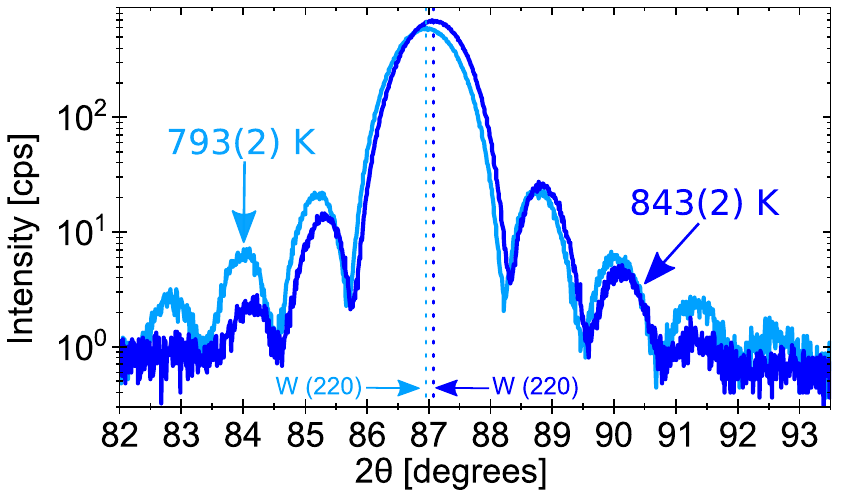}
\caption{X-ray diffraction patterns of 100~{\AA} W layers grown at 793(2) and 843(2) K on single crystalline Al$_{2}$O$_{3}$~($11\bar{2}0$) substrates. The films are capped with 30~{\AA} Al. The indexed peak relates to the $\alpha$-W structure.}
\label{fig:strain}\hfill
\end{figure}

We established that the epitaxial growth and quality of W thin films are highly dependent on substrate and thin film thickness. Following this, the dependence on the deposition temperature will now be discussed. Fig.~\ref{fig:strain} shows a diffraction pattern around the W~($220$) Bragg peak of two 100~{\AA} thick W thin films grown at different deposition temperatures on Al$_{2}$O$_{3}$~($11\bar{2}0$) substrates. With an increase in deposition temperature of 50~K, the Bragg peak position shifts slightly towards higher 2$\theta$ angles, corresponding to an atomic spacing $d_{220}$ of 1.119(0)~{\AA} and 1.118(0)~{\AA} for 793(2) and 843(2)~K, respectively. A more distinct difference is visible in the Laue oscillations around the main peak, which are visible over an angular range of 8 and 11~degrees for 843(2) and 793(2)~K, respectively.

While the Laue oscillations are symmetric for the sample grown at 793(2)~K, they show an asymmetry for the sample deposited at 843(2)~K, decaying faster on the low than on the high angle side. In contrast, epitaxial W grown on MgO~($001$) or Fe on MgAl$_2$O$_4$~($001$) \cite{Ravensburg2022} exhibit Laue oscillations around the main Bragg peaks which decay faster on the high than on the low angle side. This difference in decay is attributed to the opposite strain state and therefore, tensile instead of compressive out-of-plane strain. Hence, for W grown on Al$_{2}$O$_{3}$~($11\bar{2}0$), the out-of-plane lattice spacing is expected to increase over the layer thickness due to strain relaxation. The change in symmetry with growth temperature observed in Fig.~\ref{fig:strain} shows that the introduction of misfit dislocations for strain relaxation in W thin films is thermal energy dependent. The increase in deposition temperature by 50~K might lead to an increase in ad-atom mobility at the interface, partly preventing relaxation through the introduction of misfit dislocations. For the film grown at 843(2)~K misfit dislocations are possibly incorporated into the growing W layer at slightly larger layer thicknesses while relaxation is expected to take place at the substrate/W interface for W grown at 793(2)~K. This is in line with the smaller interplanar spacing of the sample deposited at 843(2)~K, since it corresponds to an average over the W layer thickness. The degree of the observed asymmetry in this sample is, however, smaller than expected from the large lattice mismatch for Al$_{2}$O$_{3}$~($11\bar{2}0$)/W~($110$), indicating that a large part of the strain is still released at the substrate/W interface. For W on Al$_{2}$O$_{3}$~($11\bar{2}0$) grown at 843(2)~K, the relative range of the oscillations on the faster decaying side is roughly 38~\% of the whole range of oscillations. For W~($001$) grown on MgO~($001$) with a smaller lattice mismatch of -6.5~\% and Fe~($001$) on MgAl$_2$O$_4$~($001$) with a lattice mismatch of only -0.2~\% the relative range spans 29~\% and 15~\% \cite{Ravensburg2022}, respectively. As W shows this intriguing growth temperature dependence in its crystal quality, the exact kinetic implications on the formation of dislocations at different thicknesses in the growing W layers will be part of an upcoming study.

\subsection{\label{sec:crystalStructure}Growth of tungsten and CoFe bilayers}\label{subsec5}

\begin{figure*}[t!] 
\centering
\includegraphics[width=\textwidth]{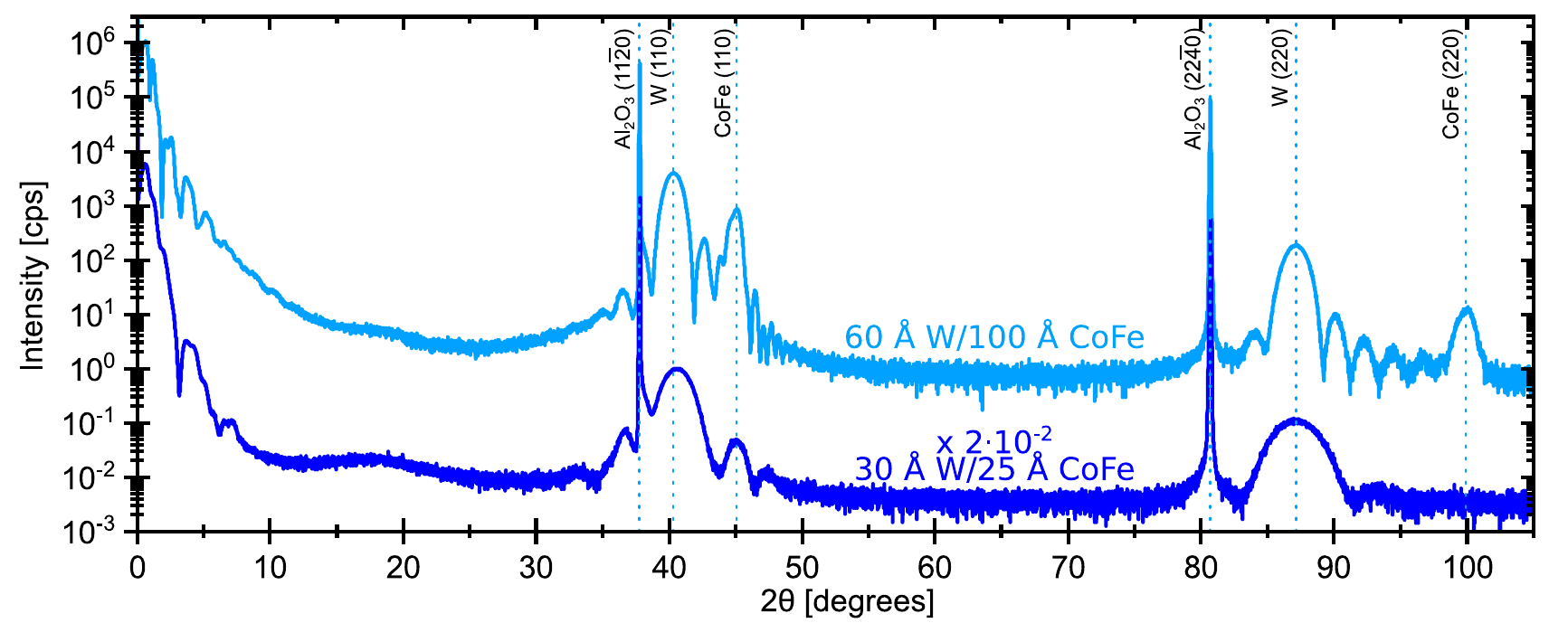}
\caption{X-ray scattering patterns of W/CoFe bilayers grown on single crystalline Al$_{2}$O$_{3}$~($11\bar{2}0$) substrates. Indexed peaks relate to the $\alpha$-W structure. The film corresponding to the upper pattern consists of 60~{\AA} W and 100~{\AA} CoFe capped with 30~{\AA} Al, while the film corresponding to the lower pattern is thinner and consists of 30~{\AA} W and 25~{\AA} CoFe capped with 60~{\AA} Al$_{2}$O$_{3}$. The lower pattern is vertically shifted for clarity.} 
\label{fig:CoFe_thickness}\hfill
\end{figure*}

Building on the knowledge about epitaxial W single layer growth, W/CoFe bilayers of different layer thicknesses have been deposited. X-ray scattering patterns of a "thick" 60~{\AA} W/100~{\AA} CoFe and a "thin" 30~{\AA} W/25~{\AA} CoFe bilayer sample both grown on Al$_{2}$O$_{3}$~($11\bar{2}0$) are displayed in Fig.~\ref{fig:CoFe_thickness}. In the small angle regime, Kiessig fringes \cite{Kiessig1931} are visible. As their spacing relates to the total film thickness including all layers, broader fringes are observed for the "thin" bilayer sample. The fringes decay at around 10 and 8~degrees in 2$\theta$ for the "thick" and "thin" bilayer sample, respectively.

In diffraction, sharp peaks corresponding to the Al$_{2}$O$_{3}$~($11\bar{2}0$) and ($22\bar{4}0$) reflections of the substrate are visible for both samples. Two peaks can be attributed to epitaxially growing W with ($110$) out-of-plane orientation, namely W~($110$) and ($220$), in line with the observations for W single layers on Al$_{2}$O$_{3}$~($11\bar{2}0$) substrates. The width of the Bragg peaks is larger as compared to the ones of the W single layers in Fig.~\ref{fig:fitting}, since the W thickness is reduced \cite{Scherrer1918}. Based on scans around off-specular reflections, the sole presence of the $\alpha$-W structure is confirmed even for the thinnest W layer in this study with a thickness of 30~{\AA}. The W peak positions relate to atomic spacings $d_{110}$ and $d_{220}$ of 2.235(0) and 1.118(0)~{\AA} for the "thick" sample and 2.221(0) and 1.118(0)~{\AA} for the "thin" bilayer sample. The change in $d_{110}$ and $d_{220}$ with W layer thickness between 30 and 1000~{\AA} is below 0.9~\%, in agreement with the described strain relaxation at the substrate/W interface and not over the W layer thickness.

For the "thick" sample, peaks corresponding to the CoFe~($110$) and ($220$) Bragg reflections \cite{Jain2013, osti_CoFe} are observed at 44.981(5) and 99.940(7)~degrees in 2$\theta$. As no other peaks in the specular 2$\theta-\theta$ scan can be attributed to CoFe, the layer is assumed to at least grow highly textured, in line with the observed FWHM of the CoFe~($220$) rocking curve of 0.06(0)~degrees. The corresponding $d_{110}$ and $d_{220}$ are 2.014(0) and 1.006(0)~{\AA}, respectively. The reported equilibrium values of CoFe are smaller, 2.007 and 1.004~{\AA} for $d_{110}$ and $d_{220}$, respectively \cite{Jain2013, osti_CoFe}. The equilibrium interplanar spacings $d_{001}$ and $d_{1\bar{1}0}$ in CoFe have a mismatch to the corresponding distances in W of approximately 12~\% each \cite{Jain2013, osti_CoFe, osti_W}. Based on its positive Poisson's ratio of 0.397 \cite{Liu1992}, CoFe~($110$) is assumed to grow with tensile in-plane and compressive out-of-plane strain on W~($110$), in line with the measured larger than equilibrium out-of-plane interplanar spacings in this study. Moreover, indications of fully epitaxial growth of parts of the CoFe layer are present in the form of Laue oscillations of two different average oscillation frequencies of approximately $f_1$~=~1.5~degrees$^{-1}$ and $f_2$~=~0.6~degrees$^{-1}$, between 35 and 48~degrees in 2$\theta$.

This oscillation frequency is directly related to the coherently scattering thickness: the oscillations of the lower frequency $f_1$ can be related to a thickness of 59~{\AA}, which is in the order of 98~\% of the W layer thickness, while the higher frequency $f_2$ oscillations relate to a thickness of around 147~{\AA}, which is in the order of 92~\% of the bilayer thickness. The observation of Laue oscillations relating to the bilayer thickness are proof of coherent scattering throughout both layers \cite{Fewster1996}. In contrast, no Laue oscillations with a frequency relating to the CoFe single layer thickness are observed around the CoFe~($110$) or ($220$) Bragg peaks. This can be attributed to multiple possible reasons: First, based on the calculation above, if 98~\% of the W layer and 92~\% of the bilayer are assumed to scatter coherently, then only 88~\% of the CoFe layer scatters coherently. As can be observed in Fig.~\ref{fig:fitting}c, the relative intensity of the Laue oscillations decreases substantially with a lower degree of coherent scattering. Second, the scattering intensity relates to the form factor and, hence, the atomic number squared $Z^2$ \cite{Fewster1996}. Therefore, the scattering intensity from W with $Z$~=~74 is expected to be higher by a factor of 7.8 compared to the scattering intensity from the alloy with $Z$~=~27 and $Z$~=~26 for Co and Fe, respectively.

\begin{figure*}[t!] 
\centering
\includegraphics[width=\textwidth]{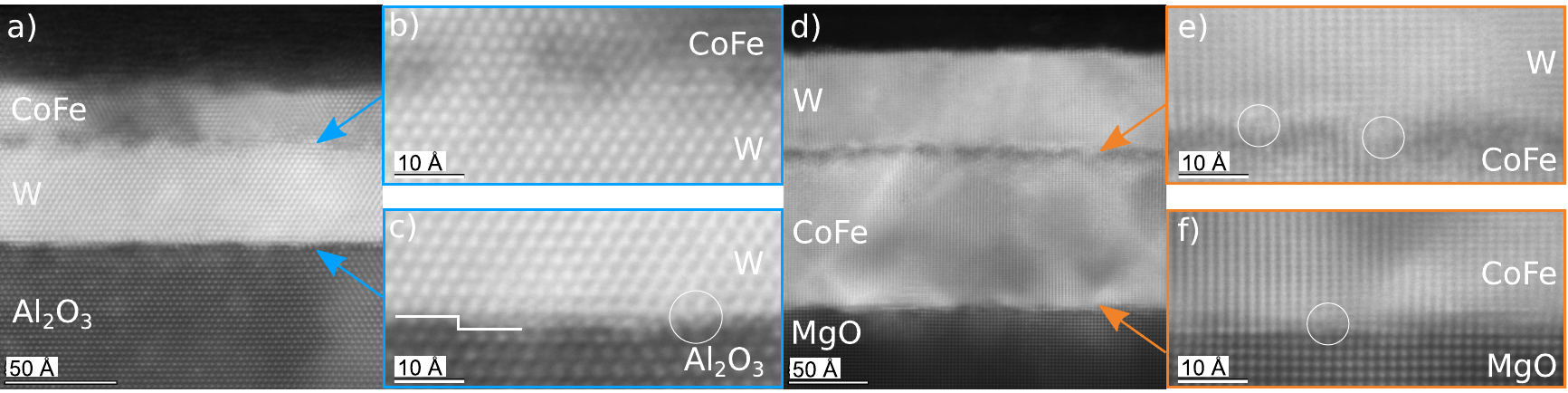}
\caption{Cross-section dark field scanning tunneling electron microscopy (STEM) images with high resolution zoom-in of a a)/b)/c) 30~{\AA} W/25~{\AA} CoFe bilayered thin film grown on a single crystalline Al$_{2}$O$_{3}$~($11\bar{2}0$) substrate capped with 60~{\AA} Al$_{2}$O$_{3}$ and a c)/d)/e) 100~{\AA} CoFe/60~{\AA} W bilayered thin film grown on a single crystalline MgO~($001$) substrate capped with 30~{\AA} Al. Visible defects at the interfaces are marked in white.} 
\label{fig:CoFe_TEM}\hfill
\end{figure*}

For the "thin" bilayer sample, only the peak at 45~degrees can be attributed to CoFe. Since its peak position overlaps with the Laue oscillations around the W~($110$) Bragg peak, a clear identification is difficult. However, the spacing between the peak's position and the Bragg reflection is different compared to the spacing between the Laue oscillations on the lower angle side and the main Bragg reflection, indicating that both do not have the same origin. The intensity of the CoFe~($220$) Bragg reflection is assumed to lie below the detection limit. However, the presence of a CoFe layer in all samples with finite thickness is confirmed by STEM imaging, results are displayed in Fig.~\ref{fig:CoFe_TEM}a and b.

\begin{figure*}[t!] 
\centering
\includegraphics[width=\textwidth]{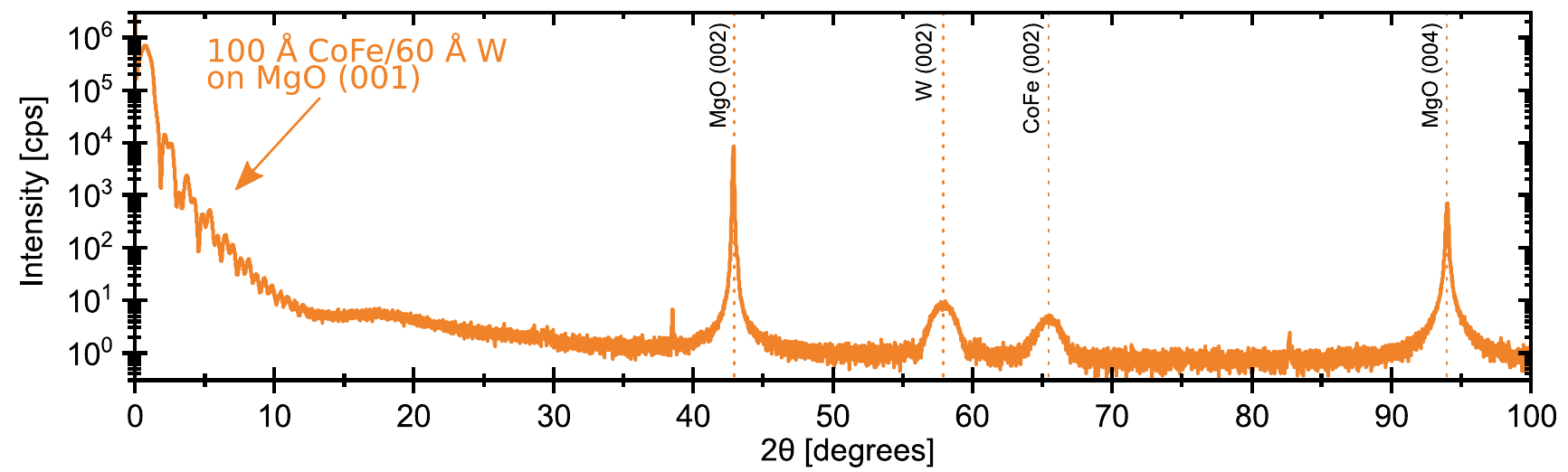}
\caption{X-ray scattering pattern of a 100~{\AA} CoFe/60~{\AA} W bilayered thin film grown on a single crystalline MgO~($001$) substrate. The film is capped with 30~{\AA} Al. Indexed peaks relate to the $\alpha$-W structure.} 
\label{fig:CoFe_substrate}\hfill
\end{figure*}

Due to the limitation of epitaxial growth being reliant on a specific substrate, just like for W, an alternative way of growing epitaxial W/CoFe bilayers was explored using a single crystalline MgO~($001$) substrate. To optimize the epitaxial growth, the lattice mismatch between MgO~($001$)/W~($001$) with 45~degrees in-plane rotation and between MgO~($001$)/CoFe were compared. The lattice mismatch between $d_{110}$ = 2.978~{\AA} of MgO~($001$) \cite{Zheng2015} and $d_{200}$ = 2.840~{\AA} of CoFe~($001$) \cite{Jain2013, osti_CoFe}) is 4.4~\%, possibly allowing for epitaxial growth with a 45~degrees in-plane rotated CoFe unit cell. A CoFe~($110$) growth orientation on MgO~($001$) is likely to be energetically less favored due to the slightly larger lattice mismatch of 4.5~\%. Both mismatches are smaller than the expected mismatch between MgO~($001$) and W~($001$) with 45~degrees in-plane rotation of -6.5~\%. Therefore, the order of the W and CoFe layers was reversed for the growth on MgO~($001$) in comparison to the bilayer grown on Al$_{2}$O$_{3}$~($11\bar{2}0$).

An x-ray scattering pattern of a 100~{\AA} CoFe/60~{\AA} W bilayer is shown in Fig.~\ref{fig:CoFe_substrate}. Pronounced Kiessig fringes \cite{Kiessig1931} are visible in the small angle regime up until 13~degrees in 2$\theta$, a larger range compared to the bilayer grown on Al$_{2}$O$_{3}$~($11\bar{2}0$). Based on the fitting of the reflectivity, the interfaces are flatter compared to the bilayer grown on Al$_{2}$O$_{3}$~($11\bar{2}0$). In diffraction, two sharp peaks are observed corresponding to ($002$) and ($004$) planes in the MgO~($001$) substrate. The CoFe~($002$) Bragg peak is observed in the specular scan at 65.452(6)~degrees, in agreement with the calculated energetically favored [$001$] growth direction on this substrate. The peak position corresponds to a $d_{002}$ = 1.425(0)~{\AA}, which is close to the chemical composition dependent equilibrium value \cite{Jain2013, osti_CoFe, Liu1992}. At 57.905(6)~degrees, a peak is observed which is attributed to W~($002$). Low intensity Laue oscillations are observed around the W~($002$) peak indicating epitaxial growth resulting in coherent scattering.

To support the results on crystal structure and epitaxial growth with a real space technique, atomic resolution cross-section STEM images were recorded for bilayers grown on both substrates. A cross-section dark-field STEM image of a W/CoFe bilayer grown on Al$_{2}$O$_{3}$~($11\bar{2}0$) is shown in Fig.~\ref{fig:CoFe_TEM}a. The image is recorded of the thin film corresponding to the lower scattering pattern displayed in Fig.~\ref{fig:CoFe_thickness}. High resolution magnifications of the substrate/W and W/CoFe interfaces are displayed in Fig.~\ref{fig:CoFe_TEM}c and b, respectively. Based on the STEM micrographs, the epitaxial growth of W and CoFe layers on Al$_{2}$O$_{3}$~($11\bar{2}0$) is confirmed. The atomically sharp boundaries indicate a single-crystalline nature for both layers. Moreover, few misfit dislocations directly at the substrate/film interface are visible as a consequence of large lattice mismatch and strain discussed in detail earlier. In addition, the rather blurry interface for a width of 1 to 2 monolayers is an indication for relaxation directly at the interface for W grown on this substrate, in agreement with the results of the x-ray scattering and DFT studies. The presence of atomic terraces with a terrace width of a few tens of {\AA}ngstr{\"o}ms for the Al$_{2}$O$_{3}$~($11\bar{2}0$) substrate is confirmed based on these images. Misfit dislocations are also observed at the W/CoFe interface, but the crystal structure of the CoFe layer appears to be single crystalline. Hence, semicoherent growth of W/CoFe on Al$_{2}$O$_{3}$~($11\bar{2}0$) is confirmed.

For comparison, a cross-section dark-field STEM image of a CoFe/W bilayer grown on MgO~($001$) is shown in Fig.~\ref{fig:CoFe_TEM}d. High resolution magnifications of the substrate/CoFe and CoFe/W interfaces are displayed in Fig.~\ref{fig:CoFe_TEM}f and e, respectively. The images are recorded of the thin film corresponding to the scattering pattern displayed in Fig.~\ref{fig:CoFe_substrate}. The STEM micrographs confirm the epitaxial growth of W/CoFe bilayers on MgO~($001$), in agreement with the x-ray scattering results. The substrate/CoFe interface seems to be sharp, with strain visible for at least the first 1 to 2 monolayers of CoFe. Dislocations are visible for the W/CoFe interface, however, the W layer itself seems to grow fully single crystalline above the interface.

\section{Conclusions}\label{sec4}

The structural properties of epitaxial $\alpha$-W thin films deposited on Al$_{2}$O$_{3}$~($11\bar{2}0$) and MgO~($001$) substrates have been studied in in-plane and out-of-plane scattering experiments as well as with real space techniques and electronic transport measurements. Emphasis was given to the overall quality of layering and crystal structure, analyzing the epitaxial relationship and growth mode in combination with {\it ab~initio} calculations. The crystal quality of W~($110$) on Al$_{2}$O$_{3}$~($11\bar{2}0$) was found to be higher compared to films of equivalent thickness on MgO~($001$) even though the lattice mismatch is larger. The improvement in the crystal quality was attributed to a semicoherent growth mode including the introduction of misfit dislocations directly and in the vicinity of the substrate/film interface, yielding nearly strain-free, highly coherent W layers for thicknesses between 30 and 1000~{\AA}. The degree of relaxation at the interface was, however, found to be highly temperature dependent. Furthermore, the epitaxial growth of W and CoFe bilayers on both substrates was found to be of high crystal quality, exhibiting coherent scattering throughout the total bilayer thickness for W/CoFe films sputtered on Al$_{2}$O$_{3}$~($11\bar{2}0$). The results of the extensive x-ray scattering analysis on the epitaxial growth were confirmed by real space high resolution STEM imaging. The detailed analysis of the growth of these epitaxial thin films contributes to an understanding of a mechanism allowing for the high crystal quality, despite a substrate/film lattice mismatch. It is of technological importance, as bilayers of W and CoFe might be essential in future spintronic applications.

\backmatter

\section*{Funding and/or Conflicts of interests/Competing interests}

\bmhead{Declarations}
The authors have no competing interests to declare.

\bmhead{Acknowledgments}
VK and PS would like to acknowledge financial support from the Swedish Research Council (Project Nos. 2019-03581 and 2021-0465). GKP acknowledges funding from the Swedish Energy Agency (Project No. 2020-005212). DM acknowledges the financial support from the Olle Engkvist Foundation (project number 217-0023). The computations were enabled by resources provided by the National Academic Infrastructure for Supercomputing in Sweden (NAISS) at National Supercomputer Centre (NSC) in Linköping partially funded by the Swedish Research Council through grant agreement no. 2022-06725.

\bmhead{Data availability}
The data that support the findings of this study are available from the authors upon reasonable request.

\bmhead{Contribution statement}
ALR, RB, GKP, PS, and VK contributed to the study conception and design. Material preparation, data collection, and analysis were performed by ALR, RB, and LS. Theoretical calculations were performed by DM. VK supervised the project. The first draft of the manuscript was written by ALR and all authors commented on previous versions of the manuscript. All authors read and approved the final manuscript.

\providecommand{\noopsort}[1]{}\providecommand{\singleletter}[1]{#1}%

\end{document}